\documentclass{article}
\usepackage{spconf,amsmath,graphicx,amssymb,makecell,caption,subcaption,makecell,hyperref}
\usepackage{multirow}
\title{Show me the instruments:\\ Musical Instrument Retrieval from Mixture Audio}
\name{\parbox{0.7\linewidth}{\centering
Kyungsu Kim\(^{1*}\),
Minju Park\(^{1*}\),
Haesun Joung\(^{1*}\),
Yunkee Chae\(^{2}\),
Yeongbeom Hong\(^{1}\),
Seonghyeon Go\(^{1}\),
Kyogu Lee\(^{1,2,3}\)}
\thanks{\(^*\)Equal contribution.}}
\address{\(^1\)Department of Intelligence and Information, Seoul National University \\
\(^2\)Interdisciplinary Program in Artificial Intelligence, Seoul National University\\
\(^3\)Artificial Intelligence Institute, Seoul National University}

\begin{document}
\ninept

\maketitle
\begin{abstract}
As digital music production has become mainstream, the selection of appropriate virtual instruments plays a crucial role in determining the quality of music.
To search the musical instrument samples or virtual instruments that make one’s desired sound, music producers use their ears to listen and compare each instrument sample in their collection, which is time-consuming and inefficient. 
In this paper, we call this task as \textit{Musical Instrument Retrieval} and propose a method for retrieving desired musical instruments using reference mixture audio as a query.
The proposed model consists of the \textit{Single-Instrument Encoder} and the \textit{Multi-Instrument Encoder}, both based on convolutional neural networks.
The Single-Instrument Encoder is trained to classify the instruments used in single-track audio, and we take its penultimate layer's activation as the instrument embedding.
The Multi-Instrument Encoder is trained to estimate multiple instrument embeddings using the instrument embeddings computed by the Single-Instrument Encoder as a set of target embeddings.
For more generalized training and realistic evaluation, we also propose a new dataset called \textit{Nlakh}.
Experimental results showed that the Single-Instrument Encoder was able to learn the mapping from the audio signal of unseen instruments to the instrument embedding space and the Multi-Instrument Encoder was able to extract multiple embeddings from the mixture audio and retrieve the desired instruments successfully.
The code used for the experiment and audio samples are available at:
\url{https://github.com/minju0821/musical\_instrument\_retrieval}
\end{abstract}
\section{Introduction}
\label{sec:intro}
Nowadays, advances in digital music production technology enabled the musicians to explore a greater range of sonic possibilities to work with.
Particularly, the development of the Digital Audio Workstation (DAW) and virtual instruments greatly expanded the space of the musical creativity~\cite{walzer2017independent}.
As there are a large number of virtual instruments with high timbral diversity and the quality of music is highly dependent on the timbre of the instruments, selecting appropriate musical instruments plays a crucial role in digital music production. 

Typical ways to retrieve proper musical instruments from a large library of instruments are listening to the audio samples of the instruments one-by-one or referring to the text description of the instruments if available.
However, listening to the audio samples is time-consuming and inefficient, and the descriptions are often unavailable or insufficient to express the subtle nuance of the timbre of the musical instruments~\cite{knees2015giantsteps}.

\begin{figure}[tb]
\setlength{\belowcaptionskip}{-7pt}
\begin{minipage}[b]{1.0\linewidth}
  \centering
  \centerline{\includegraphics[width=8.5cm]{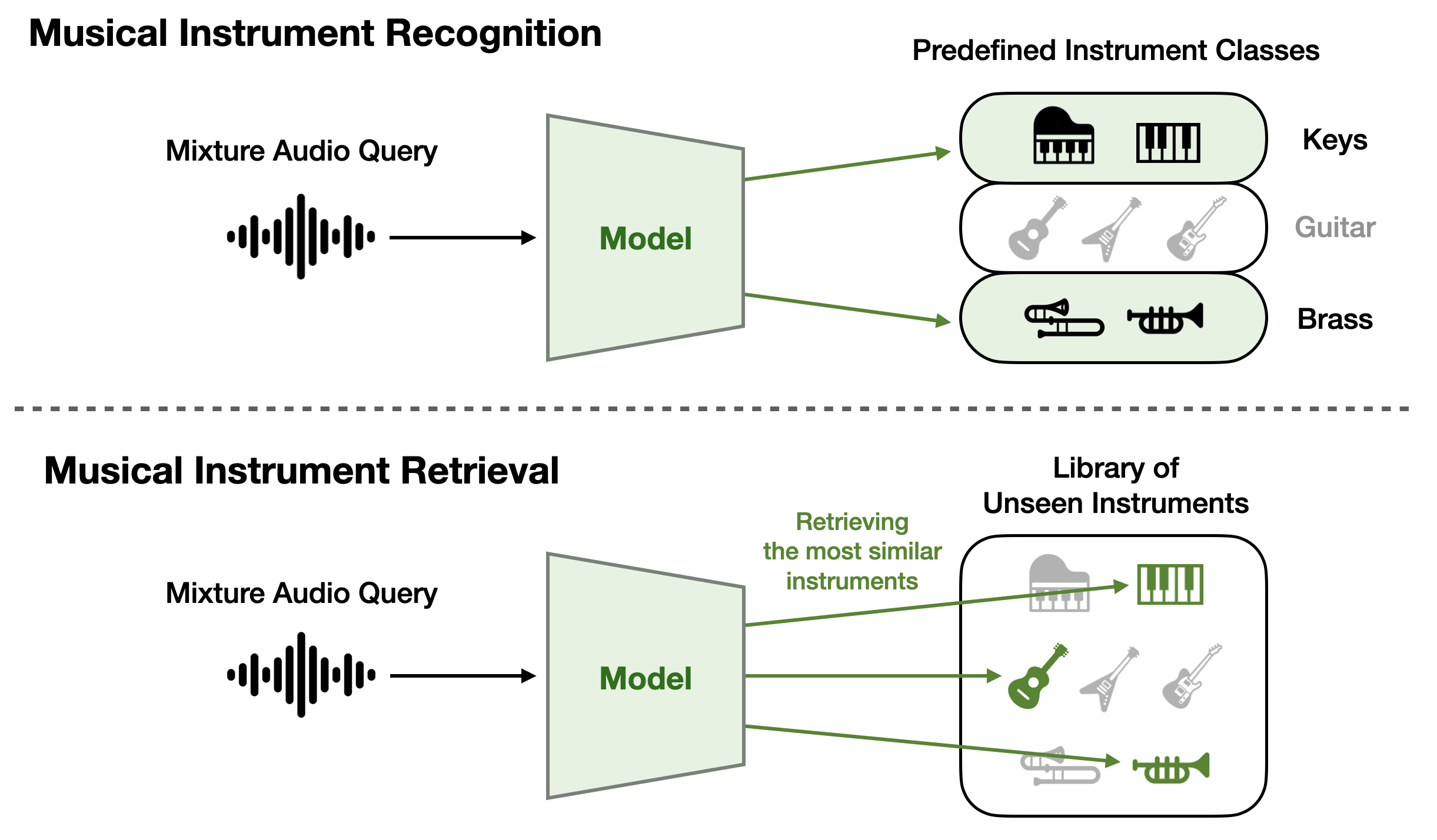}}
\end{minipage}
\caption{Comparison between musical instrument recognition and retrieval task.}
\label{fig:task}
\end{figure}

We call this task of retrieving specific desired instruments from the library of musical instruments as \textit{Musical Instrument Retrieval}.
Since musicians often refer to existing music to describe the sound they want, we propose to use reference music as a query for musical instrument retrieval.
In this task, given a mixture audio query, the model has to retrieve the instruments that most closely resemble the instrument used in the mixture audio query.
In our experiment, for quantitative evaluation, the instrument used for mixture audio query was always included in the library.
We evaluated whether the model retrieved the exact instruments used in mixture audio query in terms of F1 score and mean Average Precision (mAP). 

Musical instrument recognition is a closely related task that has been actively studied in the field of music information retrieval~\cite{han2016deep,avramidis2021deep,kratimenos2021augmentation,li2015automatic,lostanlen2016deep,cheuk2022jointist}.
However, existing methods of musical instrument recognition rule out the unique character of each instrument and only predicts the coarse categories of the instrument so that it cannot be directly used for fine-grained musical instrument retrieval task.
Comparison between the two tasks is illustrated in Fig.~\ref{fig:task}.

\begin{figure*}
    \centering
    \begin{tabular}{cc}
    \begin{subfigure}{\columnwidth}
        \centering
        \caption{Training Single-Instrument Encoder.}
        \label{fig:single_enc}
        \vspace*{0.25cm}
        \includegraphics[width=\columnwidth]{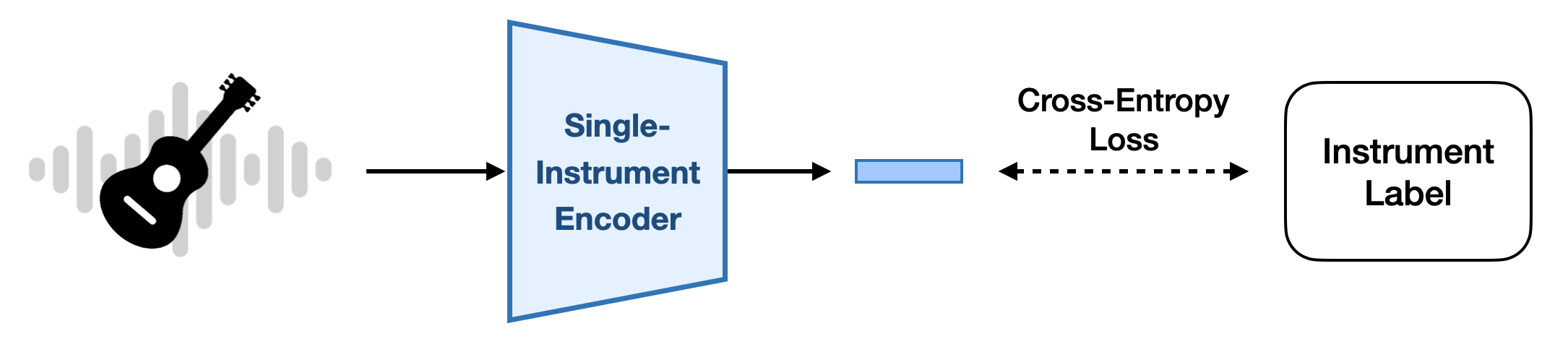}
    \end{subfigure}
    &
    \\
    \hfill
    \vspace*{0.1cm}
    \begin{subfigure}{\columnwidth}
        \centering
        \caption{Training Multi-Instrument Encoder.}
        \label{fig:multi_enc}
        \vspace*{0.25cm}
        \includegraphics[width=\columnwidth]{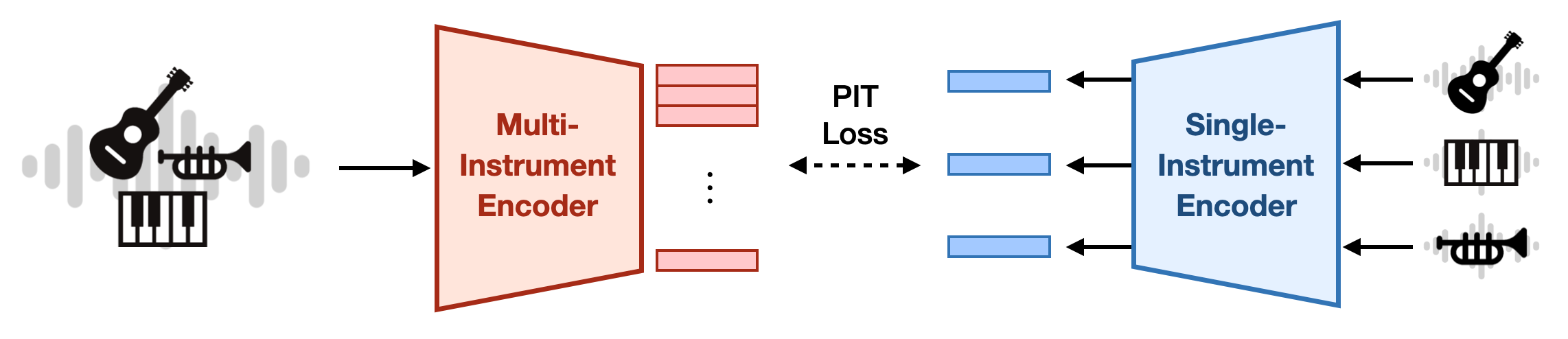}
    \end{subfigure}
    &
    \multirow[t]{2}[0]{*}[0.7cm]{
        \begin{subfigure}{\columnwidth}
        \centering
        \caption{Retrieving similar instruments from the library using proposed method.}
        \label{fig:single_enc}
        \vspace*{0.25cm}
        \includegraphics[width=\columnwidth]{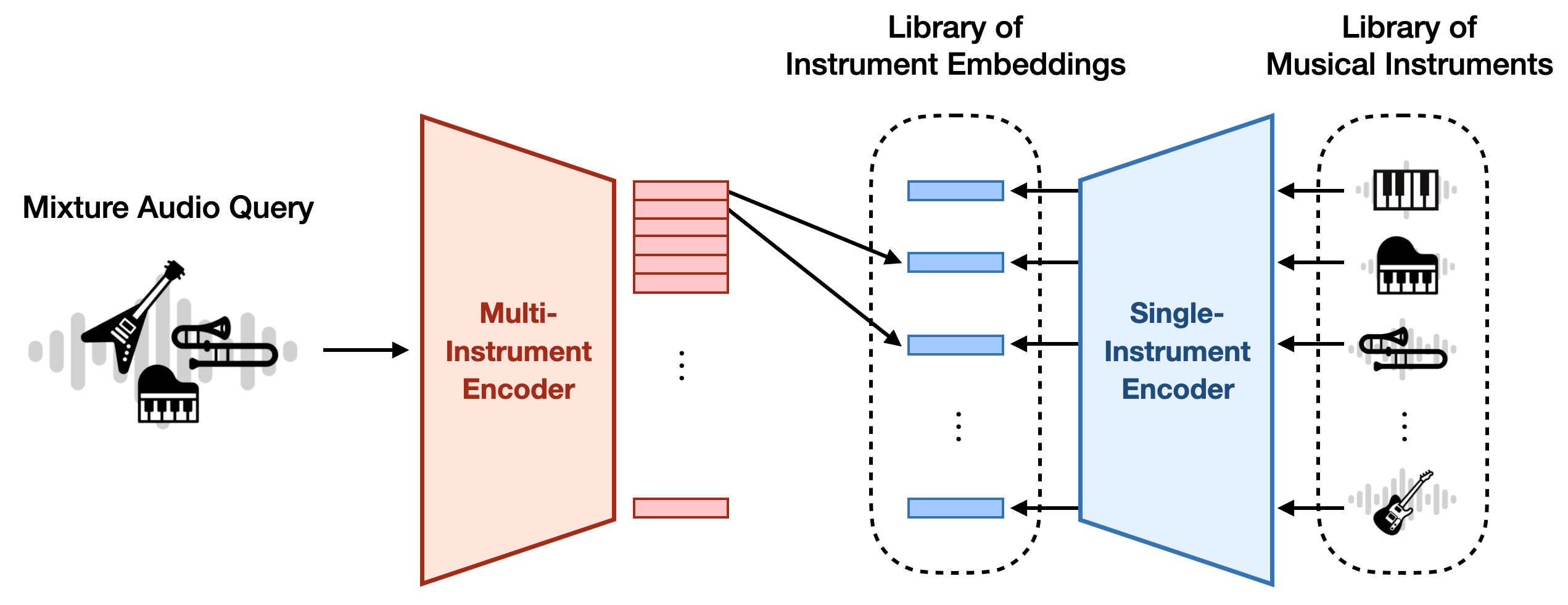}
        \end{subfigure}
    }
    \end{tabular}
    \vspace*{0.1cm}
    \caption{The overall process of the suggested method. (a) Single-Instrument Encoder is trained to classify which instrument played the input audio. We take the penultimate layer's activation of the trained network as instrument embedding. (b) Multi-Instrument Encoder extracts multiple instrument embeddings from the mixture audio. The Single-Instrument Encoder provides the set of target embeddings. (c) At inference time, we first extract the instrument embeddings of each instrument in the instrument library for a single time. Then we extract the multiple embeddings from the mixture audio query and retrieve the most similar instruments from the instrument library.}
    \label{fig:model}
\end{figure*}

Our proposed method employs the Single-Instrument Encoder and the Multi-Instrument Encoder.
The Single-Instrument Encoder extracts the instrument embedding from single-track audio.
Using the embeddings extracted by the Single-Instrument Encoder as the target, the Multi-Instrument Encoder is trained to extract multiple numbers of instrument embeddings from the mixture audio.
Since we estimate the set of embeddings, which is permutation-invariant, we use permutation invariant training (PIT)~\cite{yu2017permutation} scheme for Multi-Instrument Encoder training.

Training and evaluating a general instrument encoder requires a dataset consisting of a large number of different instruments.
At the same time, the dataset should contain ensembles of different instruments to enable the model to extract embeddings robustly to instrument combinations and performance.
To meet these conditions, we propose a new dataset called \textit{Nlakh} (pronounced as en-läk), which is a combination of the NSynth dataset~\cite{nsynth2017} and the Lakh dataset~\cite{raffel2016learning,lakh}.
Based on the implementation of ~\cite{Martel_NSynth-MIDI-Renderer_2019}, we rendered the MIDI files in the Lakh dataset by using note samples in the NSynth dataset.

Experimental results show that the Single-Instrument Encoder successfully maps the different audio samples of the same instruments into close embeddings. 
Results also show that the Multi-Instrument Encoder is able to separate the mixture audio at the embedding level and retrieve desired instruments from the library successfully.

\section{Related Works}
\label{sec:relatedworks}

Studies on retrieving musical instruments or extracting instrument embeddings are still in its early stages. 
Recently,~\cite{shi2022use} has trained and evaluated a model for extracting instrument embedding from a music signal by adopting the framework of speaker verification task, but the model was limited to extracting an embedding from single-sourced audio.
Musical instrument retrieval methods with audio query have also been studied recently, but mostly focusing on retrieving drum samples. 
~\cite{mehrabi2018similarity} adopts deep convolutional auto-encoders to retrieve drum samples by using vocal imitations as the query.
Furthermore, ~\cite{kim2020drum} conducts deep metric learning to extract the drum embeddings from a mixture audio as the query.
In this paper, we expand this approach for retrieving multi-pitched instruments.

\section{Method}
\label{sec:pagestyle}

The proposed model consists of the Single-Instrument Encoder and the Multi-Instrument Encoder.
The Single-Instrument Encoder extracts an instrument embedding from a single-track audio of the instrument.
Using the instrument embeddings computed by the Single-Instrument Encoder as a set of target embeddings, the Multi-Instrument Encoder is trained to estimate the multiple instrument embeddings.
As we estimate the set of embeddings, which is permutation-invariant, PIT scheme~\cite{yu2017permutation} was used for training.
The overall framework of the proposed model is depicted in Fig.~\ref{fig:model}.

\subsection{Single-Instrument Encoder}
\label{ssec:subhead_f}

In order to extract an instrument embedding from single-track audio with the Single-Instrument Encoder, we trained a network performing classification to match the audio samples with their instrument labels.
We used the network's penultimate layer's activation as the instrument embedding, which is a 1024-dimensional vector.
For an instrument $i_k$, the Single-Instrument Encoder $f$ extracts the embedding of the instrument $i_k$ as $f(x_{i_k})$, where $x_{i_k}$ is the single-track audio of the instrument $i_k$.

\subsection{Multi-Instrument Encoder}
\label{ssec:subhead}

The Multi-Instrument Encoder $g$ aims to estimate the embeddings of a set of instruments $I=\{{i_1, i_2,...,i_N}\}$ given a mixture audio $m=\sum_{i\in I} x_{i}$. 
The target embeddings are the outputs of the Single-Instrument Encoder.
We designed the Multi-Instrument Encoder to output $M$ possible embeddings, where $M$ was set as the maximum number of instruments in a mixture audio in the training set.

The Multi-Instrument Encoder $g$ is trained to minimize the cosine embedding loss between the optimal permutation of the set of output embeddings $G=\{ g(m)_{1,:}, g(m)_{2,:}, ..., g(m)_{M,:}\}$ and the set of target embeddings $F=\{ f(x_1), f(x_2), ..., f(x_N)\}$.
To compensate for the difference in the number of embeddings and the indeterminacy of the instrument order, we used the idea of permutation invariant training to compute the loss function~\cite{yu2017permutation}.
The minimized loss function is described as follows:
\begin{align*}
    \mathcal{L} &= \min\limits_{\pi} \sum \limits_{n=1}^N \big( 1 - \cos\theta_{\pi(n),n} \big) \\
    \cos \theta_{\pi(n),n} &= \frac{g(m)_{{\pi(n)},:} \cdot f(x_n)}{||g(m)_{{\pi(n)},:}|| \cdot ||f(x_n)||}
\end{align*}
where $\pi:\{1,2,\dots,N\} \mapsto \{1,2,\dots,M\}$ is an injective function.

To minimize the computational cost of finding the optimal permutation, we applied the optimal permutation invariant training method that utilizes the Hungarian algorithm~\cite{dovrat2021many, kuhn1955hungarian}.

\subsection{Inference}
\label{section:inference}
To use the trained encoders for retrieval task, for each instrument $l_k$ in instrument library $L = \{ l_1, l_2, l_3, ..., l_K\}$, we extract the instrument embedding $f(x_{l_k})$ to construct the embedding library $E = \{f(x_{l_1}), f(x_{l_2}), ..., f(x_{l_K}) \} $ using the trained Single-Instrument Encoder.
Given the mixture audio query $m$, we extract output embeddings $\{g(m)_{1,:}, ..., g(m)_{M,:}\}$ using the trained Multi-Instrument Encoder.
Then we calculate the cosine similarity $\cos \phi_{j,k}$ as follows.
\begin{align*}
    \cos \phi_{j,k} &= \frac{g(m)_{j,:} \cdot f(x_{l_k})}{||g(m)_{j,:}|| \cdot ||f(x_{l_k})||}
\end{align*}

For each output embedding $g(m)_{j,:}$, we pick the instrument $l_k$ whose cosine similarity $\cos \phi_{j,k}$ is the largest among other instruments in $L$.
Therefore, the set of retrieved instruments $R$ given mixture audio query $m$ can be formulated as follows.
\begin{align*}
    R &= \{ l_{k'} | k' \in \{\operatorname*{argmax}_k \cos \phi_{j,k}\}_{j=1}^{M} \}
\end{align*}
Note that more than two output embeddings may be assigned to the same instrument.
Therefore, the size of a set $R$ may be smaller than $M$.

\begin{figure}[tb]
    \centering
    \begin{subfigure}[b]{\columnwidth}
        \centering
        \includegraphics[width=\columnwidth]{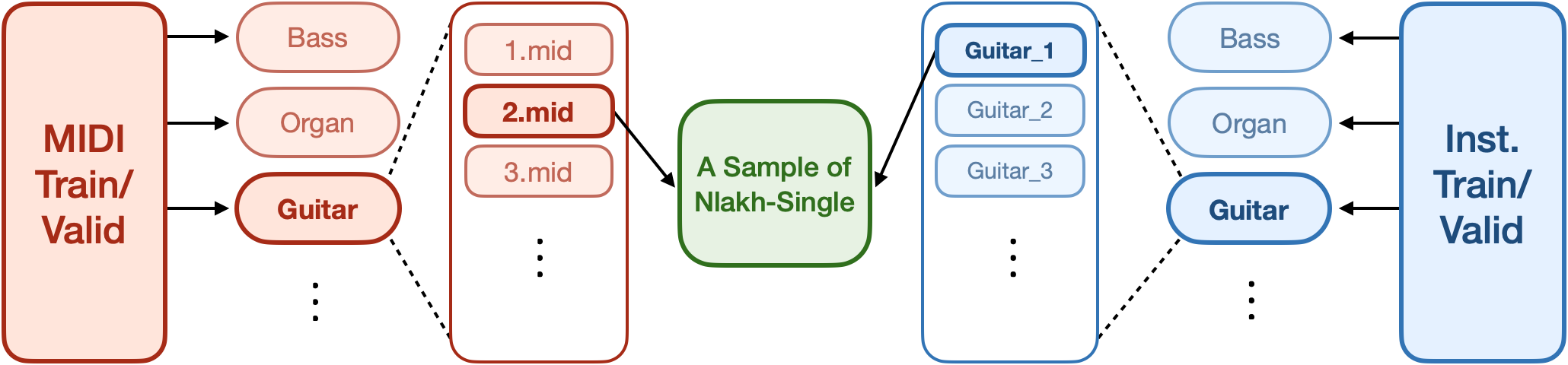}
        \caption{}
        \label{fig:data1}
    \end{subfigure}
    \hfill
    \begin{subfigure}[b]{\columnwidth}
        \centering
        \includegraphics[width=\columnwidth]{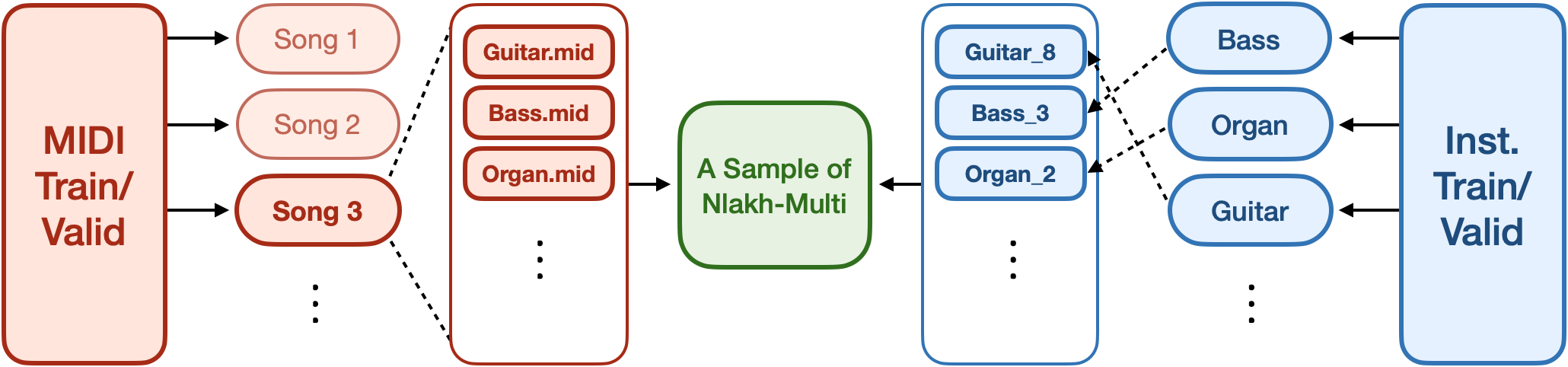}
        \caption{}
        \label{fig:data2}
    \end{subfigure}
    \caption{The process of rendering a sample of (a) Nlakh-single and (b) Nlakh-multi }
    \label{fig:data}
\end{figure}

\begin{table}[tb]
\centering
\hfill
\resizebox{\columnwidth}{!}{
\begin{tabular}{lccc}
\Xhline{2\arrayrulewidth}
\textbf{Dataset} & \begin{tabular}[x]{@{}c@{}} Size \\ (Hours)  \end{tabular} & \begin{tabular}[x]{@{}c@{}} Number of \\  Instruments \\ (Categories) \end{tabular} & \begin{tabular}[x]{@{}c@{}} Stem \\ Availability\end{tabular} \\ \hline 
Nlakh-single (ours) & 1,397  & 1,006  & $\checkmark$ \\
Nlakh-multi (ours) & 153 & 1,006 & $\checkmark$ \\ \hline
Slakh~\cite{manilow2019cutting} & 145 & 158 & $\checkmark$ \\
MUSDB18~\cite{rafii2017musdb18} & 10 & (5) & $\checkmark$ \\
MedleyDB~\cite{bittner2014medleydb} & 7 & (80) & $\checkmark$ \\
OpenMIC~\cite{humphrey2018openmic} & 56 & (20) & - \\
IRMAS~\cite{bosch2012comparison} & 6 & (11) & - \\
\Xhline{2\arrayrulewidth}
\end{tabular}
}
\caption{Comparison with other datasets.}
\label{tab:dataset}
\end{table}

\section{The Nlakh Dataset}
\label{ssec:subhead}
To train and evaluate the proposed model, the dataset should have a large number of different instruments. 
Also, the dataset should contain the ensembles of different instruments to enable the model to extract instrument embeddings robustly to instrument combinations and performance. 
However, no existing dataset fully met these requirements. Therefore, we propose a new dataset called \textit{Nlakh} that combines the NSynth dataset, which provides a large number of instruments, and the Lakh dataset, which provides multi-track MIDI data.

Nlakh consists of \textit{Nlakh-single} that contains single-track audio and \textit{Nlakh-multi} that contains mixture audio with separate tracks (stem) of each instrument.
To make Nlakh-single, we first separated each MIDI track of the Lakh dataset and categorized the tracks by their instrument family (bass, organ, guitar, etc.) according to the MIDI program number.
Then for each instrument of NSynth, we randomly selected a five-second-long excerpt from MIDI tracks in the corresponding instrument family.
For example, if the selected instrument's family is the guitar, only the MIDI files in the guitar category are used for rendering.
We rendered 1,000 samples for each instrument. 
In total, there are 1,006,000 samples in Nlakh-single.
Nlakh-single is split into train/valid set following the instrument split of NSynth (953/53).

To make Nlakh-multi, we first find a five-second-long multi-track MIDI section containing at least two valid tracks in which at least three notes are played.
Likewise in Nlakh-single, we randomly selected instruments for rendering the multi-track MIDI excerpt within the corresponding instrument family.
The Nlakh-multi has 100,000 samples for the training set and 10,000 samples for the validation set.
The overall process of making the dataset is illustrated in Fig.~\ref{fig:data}.

Among other multi-track music datasets that contains audio data, to the best of our knowledge, Nlakh has the largest number of instruments and the largest amount of data at the same time (Table~\ref{tab:dataset}).
In addition to the rendered audio dataset, we also provide a codebase to generate our dataset, so one can use it to render more samples.

\begin{table*}[t]
\centering
\parbox{11cm}{\caption{Performance of the Multi-Instrument Encoder. Small/Large indicates the size of the model. Nlakh/Random indicates which dataset is used for training.}
\label{tab:g_eval}
}
\begin{tabular}{@{\extracolsep{4pt}}lcccccc@{}}
\Xhline{2\arrayrulewidth}
\multirow{3}{*}{\textbf{Model}} & \multicolumn{2}{c}{\textbf{Family}} & \multicolumn{4}{c}{\textbf{Instrument}} \\
\cline{2-3} \cline{4-7}
& \multicolumn{2}{c}{\textbf{F1}} & \multicolumn{2}{c}{\textbf{F1}} & \multicolumn{2}{c}{\textbf{mAP}} \\ \cline{2-3} \cline{4-5} \cline{6-7}
& macro & weighted & macro & weighted & macro & weighted \\ \hline
Chance & 0.343 & 0.437 & 0.065 & 0.077 & - & -\\ \hline
Small-Nlakh & 0.626 & 0.723 & 0.482 & 0.524 & 0.553 & 0.597 \\
Large-Nlakh & 0.640 & 0.728 & 0.533 & 0.578 & 0.635 & 0.666 \\ 
Small-Random & 0.691 & 0.697 & 0.528 & 0.543 & 0.598 & 0.615 \\ 
Large-Random & \textbf{0.814} & \textbf{0.817} & \textbf{0.694} & \textbf{0.712} & \textbf{0.752} & \textbf{0.760} \\ 

\Xhline{2\arrayrulewidth}
\end{tabular}
\end{table*}

\section{Experiments}

\subsection{Single-Instrument Encoder}

We used the convolutional neural network architecture that was used in~\cite{han2016deep} for the instrument recognition task as the backbone network of the Single-Instrument Encoder, using mel-spectrogram of the audio as the input.
We used Adam optimizer with a learning rate of 0.001, and set batch size as 32.

To evaluate the Single-Instrument Encoder, we adopted the method proposed by ~\cite{shi2022use}, which used automatic speaker verification evaluation methodologies for evaluating the instrument embeddings.
We first extract the embeddings of five different samples of the target instrument by using the trained Single-Instrument Encoder. 
The average of those embeddings is used as enrollment embedding.
We also make a comparison set that contains 20 embeddings from the target instrument and 20 embeddings from the other instruments.
Then we compare each embedding in the comparison set with the enrollment embedding in terms of cosine similarity.
Verifying whether the embeddings in the comparison set correspond to the enrollment embedding or not, we compute the false reject rate and false accept rate for each instrument.
We computed the average value of equal error rate (EER), which describes the point where the false reject rate and false accept rate are equal.

\begin{figure}[t]
    \centering
    \begin{subfigure}[b]{0.49\columnwidth}
        \centering
        \includegraphics[width=\columnwidth]{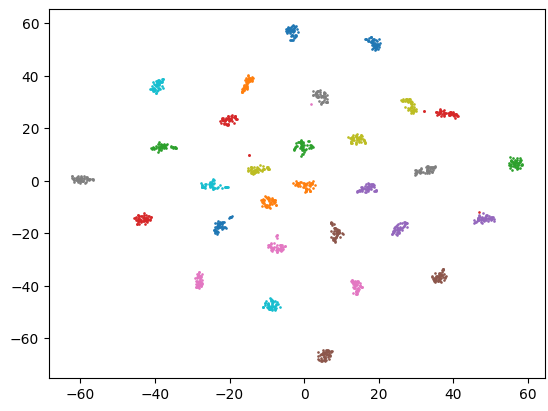}
        \caption{}
        \label{fig:data1}
    \end{subfigure}
    \hfill
    \begin{subfigure}[b]{0.49\columnwidth}
        \centering
        \includegraphics[width=\columnwidth]{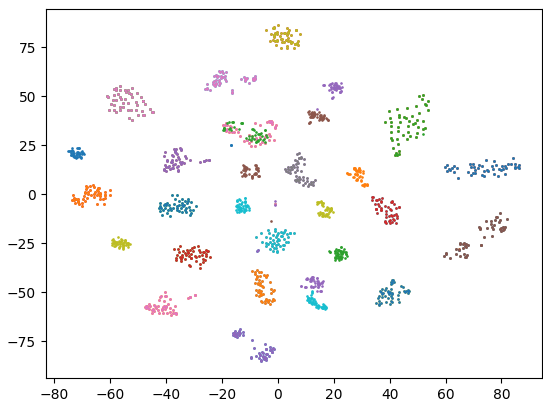}
        \caption{}
        \label{fig:data2}
    \end{subfigure}
    \caption{The t-SNE results of Single-Instrument Encoder on Nlakh-single (a) training and (b) validation dataset.}
    \label{fig:tsne_single}
\end{figure}

The average EER of our model on Nlakh-single was 0.026 while the previous work's EER on the NSynth dataset was 0.031.
Note that the samples of the NSynth dataset contain only a single note, while the samples of Nlakh-single contain multiple notes.
We also visualized the instrument embeddings of training set and validation set using t-distributed stochastic neighbor embedding (t-SNE)~\cite{van2008visualizing} in Figure~\ref{fig:tsne_single}.
The results show that the Single-Instrument Encoder could cluster the instrument embeddings robustly even for the unseen instruments in the validation set.

\subsection{Multi-Instrument Encoder}

The Multi-Instrument Encoder is trained to extract the embedding of each instrument in the mixture audio.
In this experiment, the Multi-Instrument Encoder extracts nine embeddings, which is the maximum number of instruments composing a single mixture in Nlakh-multi. 
For the network architecture of the Multi-Instrument Encoder, we tried two different network architectures.
The first is the same architecture~\cite{han2016deep} as the Single-Instrument Encoder.
We also tried a larger convolutional neural network~\cite{liu2022convnet} since the task for the Multi-Instrument Encoder is more difficult than that of the Single-Instrument Encoder.
We used Adam optimizer with a learning rate of 0.001 and set batch size as 128 for all cases.

During the experiment, we noticed an imbalance of the instrument distribution in Nlakh-multi, which may harm the performance of the trained network.
To solve this issue, we also trained the network with randomly-mixed audio.
We randomly selected a number of musical instruments between two and nine, and then randomly picked the audio samples of selected instruments from Nlakh-single.
Those samples were used to mix the randomly-mixed audio.
Rather than rendering the finite set of samples for the randomly-mixed dataset, we mixed the audio on-the-fly during training.

Given mixture audio query, we retrieved the instruments as described in Section~\ref{section:inference} and computed F1 score.
We also calculated the F1 score with the instrument family as the basis of the evaluation.
The instrument family is a coarsely categorized class of instruments, which is predefined in NSynth dataset.
To calculate the mean Average Precision (mAP), we used the highest cosine similarity between the output embeddings and each embeddings in the embedding library as the similarity score.

Table \ref{tab:g_eval} shows the evaluation results of the Multi-Instrument Encoder. We had three main observations from the evaluation.
First, every trained network performed significantly better than the chance level in all measurements.
Second, the network trained with randomly-mixed audio showed less overfitting than the network trained with Nlakh-multi.
Third, the network using the larger convolutional neural network showed better performance.
The larger convolutional neural network learns more general information and therefore can better handle the extraction of the embedding from an input mixture audio.

\section{Conclusion}
In this work, we proposed a novel method for musical instrument retrieval that employs the Single-Instrument Encoder and the Multi-Instrument Encoder to extract the instrument embeddings.
To train and evaluate the proposed model, we suggested the Nlakh dataset, which contains single-track audio and mixture audio from a large number of different musical instruments.
The evaluation result showed that the Single-Instrument Encoder was able to learn the mapping from the audio signal of unseen instruments to the instrument embedding space, and the Multi-Instrument Encoder was able to extract multiple embeddings from the mixture audio and retrieve the desired instruments successfully.
In the future, we plan to improve the robustness of our method by elaborating our dataset with appliance of various audio effects and expansion of the instrument classes.

% \section{COPYRIGHT FORMS}
% \label{sec:copyright}

% You must submit your fully completed, signed IEEE electronic copyright release
% form when you submit your paper. We {\bf must} have this form before your paper
% can be published in the proceedings.

% \vfill\pagebreak

% \section{REFERENCES}
% \label{sec:refs}

\vfill\pagebreak
\bibliographystyle{IEEEbib}
\bibliography{strings,refs}

\end{document}